\documentclass[usenatbib]{mn2e}
\usepackage{natbib}
\usepackage{url}
\usepackage{rotating}
\usepackage{caption}
\usepackage{subcaption}
\usepackage{amsmath}
\usepackage{amsfonts}
\usepackage{amssymb}
\usepackage{graphicx}
\usepackage{float}
\usepackage{color}
\usepackage{hyperref}

\hypersetup{
    pdfnewwindow=true,      % links in new window
    colorlinks=true,        % false: boxed links; true: colored links
    linkcolor=blue,         % color of internal links
    citecolor=black,        % color of links to bibliography
    filecolor=blue,         % color of file links
    urlcolor=blue           % color of external links
}

\providecommand{\apj}[0]{ApJ}
\providecommand{\apjl}[0]{ApJ Lett.}

\providecommand{\mnras}[0]{MNRAS}

\providecommand{\nat}[0]{Nature}

\begin{document}

\title[Constraining the Jet Structure of GRBs]{Constraining the Jet Structure of Gamma-Ray Bursts from Viewing Angle Observations}

\author[N. Miller, S. M\'arka, I. Bartos]{N. Miller, S.~M\'arka, I.~Bartos\thanks{ibartos@phys.columbia.edu}\\
Department of Physics, Columbia University, New York, NY 10027, USA}

\maketitle

\begin{abstract}
The angular dependence of emission in gamma-ray bursts (GRB) is of fundamental importance in understanding the underlying physical mechanisms, as well as in multimessenger search efforts. We examine the prospects of using reconstructed GRB jet opening angles and off-axis observer angles in determining the jet structure. We show that the reconstructed angles by \cite{ryan2015gamma} are inconsistent with uniform jet structure. We further calculate the number of GRBs with accurately reconstructed opening and observer angles necessary to differentiate between some phenomenological non-uniform structures.
\end{abstract}


\section{Introduction}

Gamma-ray bursts (GRB) are produced by relativistic jets, making the emission highly collimated. The specific angular profile of emission, however, is currently not understood \citep{2007RMxAC..27..140G}. A primary constraint on the jet structure, the presence of beaming, can often be inferred from the GRB afterglow observations \citep{1999ApJ...519L..17S}. Further information about the structure, however, is more difficult since we only observe each GRB from a specific angle.

There is reason to believe that the brightness profile of GRBs is structured. In particular, afterglow observations for some GRBs indicate a two component jet with an inner, narrow ultra-relativistic component, and an outer, wider mildly relativistic part \citep{2000ApJ...538L.129F,2003Natur.426..154B,2003ApJ...595L..33S,2005MNRAS.360..305S}. Direct probes of jet structure using the prompt emission at this point remain inconclusive (\citealt{pescalli2015luminosity}; and references therein).

Different theoretical models have been proposed that suggest structured emission. \cite{lipunov2001gamma} proposes universal emission profile identical for all GRBs that is consistent with the observed GRB energy distribution. Such a universal profile could explain the observed GRB energies without requiring an artificial distribution for GRB opening angles. \cite{zhang2002gamma} Suggests a similar, structured energy distribution that can explain the achromatic break time in broadband afterglow light curves via the expected viewing angle distribution, not requiring an artificial distribution of opening angles. These authors, along with other studies such as that of \cite{pescalli2015luminosity} that use the observed GRB luminosity function, suggest GRB energy distribution, $E_{\gamma}$, should vary with the observer's off-axis viewing angle $\theta_{obs}$ as
\begin{equation}
E_{\gamma} \propto \theta_{obs}^{-s}
\end{equation}
where $s$ is the structure index. As both short and long GRBs arise from broadly similar conditions, this distribution should be applicable to both categories. \cite{zhang2002gamma} and \cite{salafia2015gamma} also consider a Gaussian energy distribution, implying a smoother transition between a brighter inner region and farther off-axis.

\begin{figure}
\centering
\includegraphics[width=0.47\textwidth]{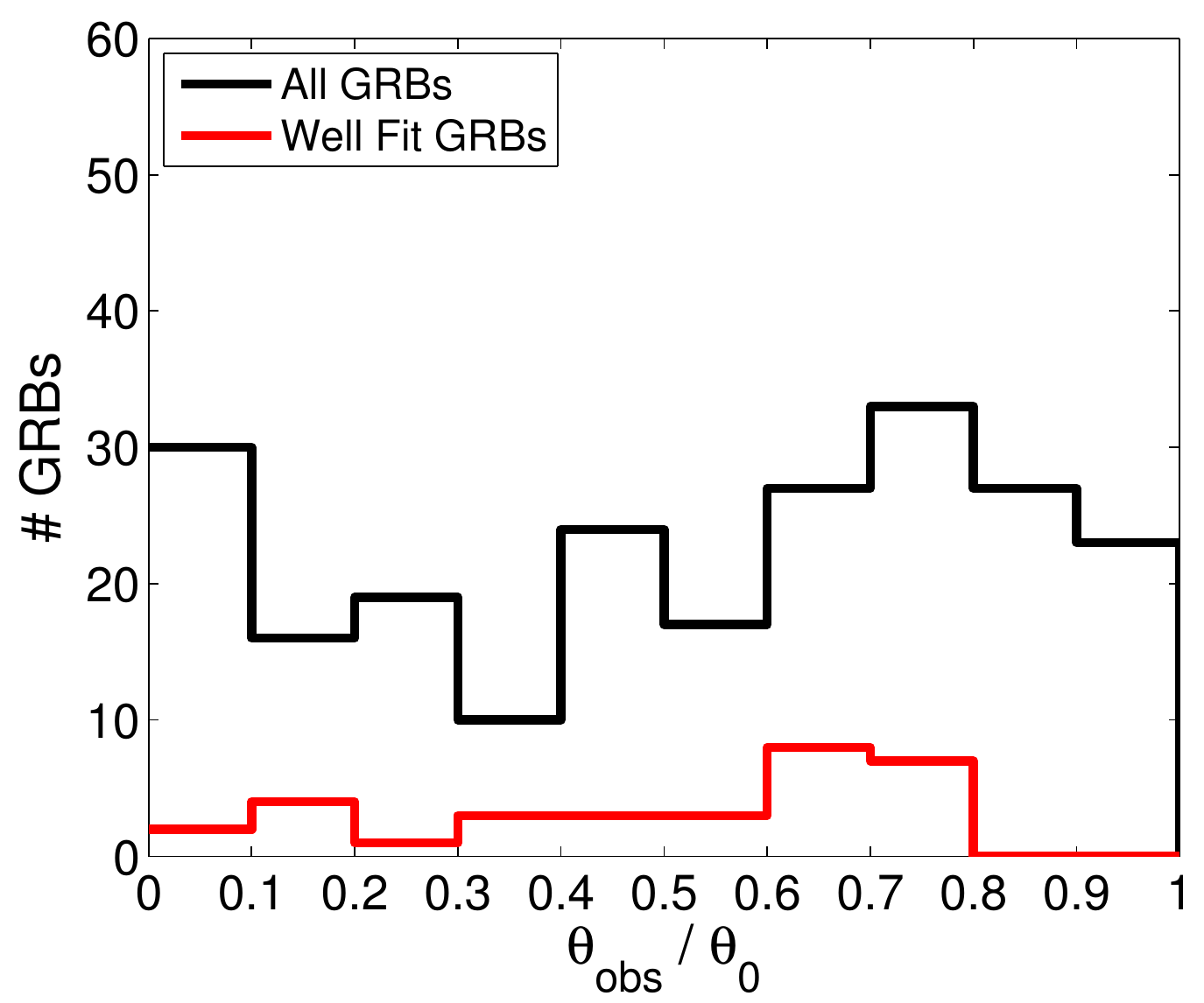}
\caption{The measured values of $\theta_{obs}/\theta_0$. ``All GRBs'' contain all 226 studied events, and ``Well Fit'' GRBs refer to events which were well-modeled by the methods of \protect\cite{ryan2015gamma}. This is a recreation of Figure 5 from \protect\cite{ryan2015gamma}, using the value of $\theta_{obs}/\theta_0$ that best fit the light curve, rather than the median value, thereby removing the characteristic peak in the center of the distribution.}
\label{vanEerten}
\end{figure}

Recently, \cite{ryan2015gamma} fit observed afterglow light curves to hydrodynamic simulations to recover jet opening angles along with, for the first time, viewing angles for $\sim 200$ afterglows observed with Swift-XRT. Their results are summarized in Fig. \ref{vanEerten}. Their simulation calculates the time evolution of spectral parameters in the afterglow, instead of relying only the time of the jet break, as has previously been the standard. The recovery of the opening and viewing angles enables a detailed study of the jet structure.

This paper examines the utilization of recovered opening and viewing angles for a set of GRBs in determining the jet structure. First, we consider the results of \cite{ryan2015gamma}, namely their ``best fit'' values for viewing angles, to determine whether they constrain jet structure models. Second, we approach the problem in general, and determine the number of GRBs with well-reconstructed viewing and observing angles needed to differentiate between representative structure models.

\section{Methods}
\label{section:methods}

For the remainder of this paper, we will use the variable $\theta = \theta_{obs}/\theta_0$, where $\theta_0$ is the jet's half opening angle. This implies $0 \leq \theta \leq 1$. Let $P(\theta, r)$ be the probability density of a GRB occuring at an angle $\theta$ and a distance $r$. Let $F (\theta,r)$ be the observed fluence of a GRB at distance $r$, seen at angle $\theta$.

Every experiment will have some minimum detectable fluence for an event, $F_{min}$. This implies that the the maximum distance at which a GRB can be observed with $\theta$ angle, $r_{max}(\theta)$, satisfies
\begin{equation}
F(\theta,r_{max}) = F_{min}
\end{equation}
The probability density of observed events, $P^*(\theta)$, will not be equal to the probability density of \textit{all} events, $P(\theta)$. This is because events can be observed only if they are brighter than $F_{min}$. The observed probability density can be written as
\begin{equation}
P^*(\theta) \propto \int_{0}^{r_{max}(\theta)} P(\theta, r) dr
\end{equation}

For uniform spatial GRB distribution, neglecting the higher-order effects of redshift (which replaces the $r^2$ term to $r^2 (1 + \lambda r)$) we have the scaling relations
\begin{equation}
P(\theta, r) \propto \theta r^2
\end{equation}
\begin{equation}
F (\theta,r) \propto E_\gamma(\theta) r^{-2}
\end{equation}
Using these relations, we obtain
\begin{equation}
P^*(\theta) \propto \int_{0}^{\big[\frac{E_\gamma(\theta)}{F_{min}}\big]^{1/2}} \theta r^2 dr \propto \theta \Big[\frac{E_\gamma(\theta)}{F_{min}}\Big]^{3/2}
\end{equation}
This allows the conversion of the observed $P^*(\theta)$ distribution into jet structure:
\begin{equation}
E_\gamma(\theta) \propto \Big[\frac{P^*(\theta)}{\theta}\Big]^{2/3}
\label{eq:ePscaling}
\end{equation}

We will use Eq. \ref{eq:ePscaling} to compare the observed $P^*(\theta)$ with different $E_\gamma(\theta)$ models.
We will assume that all GRBs can be, to reasonable accuracy, represented with one model and one set of parameters.
We will consider three jet structures: a \emph{uniform} energy distribution:
\begin{equation}
E_\gamma(\theta) = \mathrm{const.}
\end{equation}
a smooth transition represented by a \emph{Gaussian} distribution:
\begin{equation}
E_\gamma(\theta) \propto \exp(-\frac{\theta^2}{2 \sigma^2}),
\end{equation}
and a quasi \emph{universal} distribution:
\begin{equation}
E_\gamma(\theta) \propto
\begin{cases}
1 & \theta \leq \theta_{cut} \\
(\frac{\theta}{\theta_{cut}})^{-s} & \theta > \theta_{cut}.
\end{cases}
\end{equation}
These three models cover the major representative possibilities considered (c.f. \citealt{2007RMxAC..27..140G}). It is worth noting that, when the probability distributions are normalized, the uniform emission model has no free parameters, the Gaussian model has one, while the universal model has two.

Note that the opening angle for the structured cases is a nominal value, i.e. it is not a strict cutoff for the emission profile.

\begin{figure*}
\centering
\includegraphics[width=0.6\textwidth]{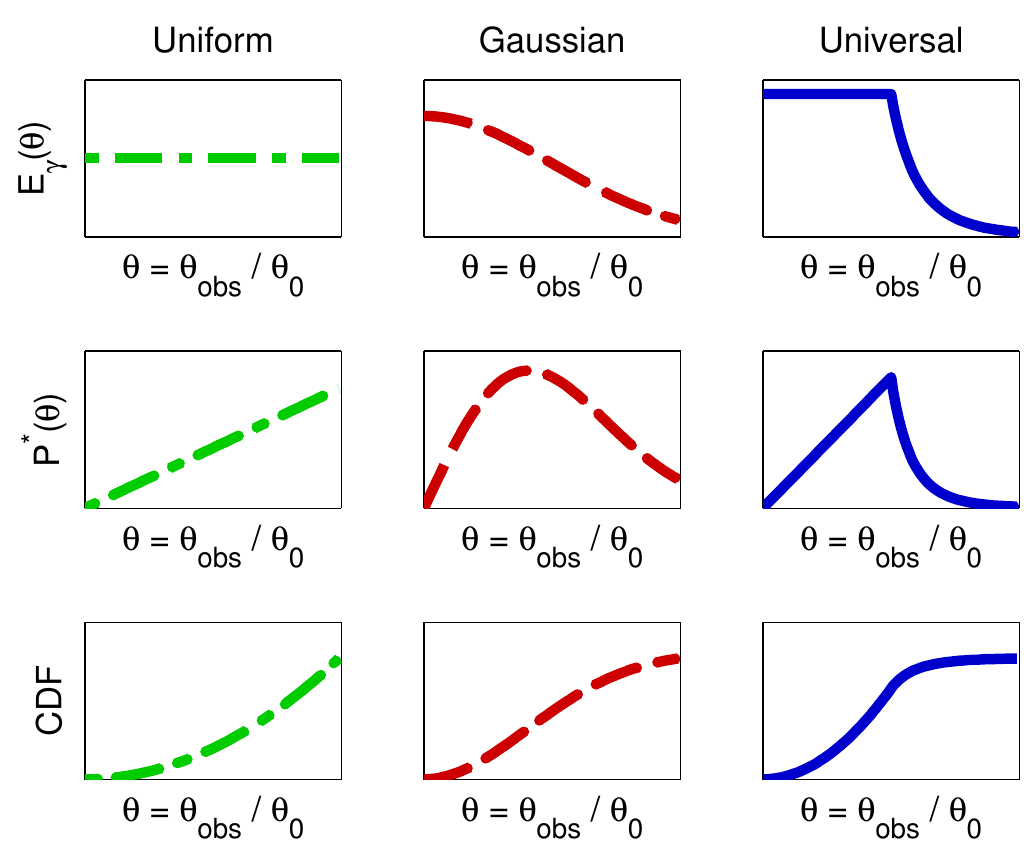}
\caption{Energy distribution (top), observed probability distribution (middle) and cumulative probability distribution (bottom) for the three representative jet structure model considered in this paper.}
\label{fig:emissionprofiles}
\end{figure*}

To probe our models, we take the cumulative distribution function (CDF) of $P^*(\theta)$ from the data obtained by \cite{ryan2015gamma}. We fit the CDF obtained using $E_\gamma(\theta)$ for the three considered models to the data. In order to determine if a CDF properly describes experimental data, we will use the p value given by the one-sample Kolmogorov-Smirnoff test.

\cite{ryan2015gamma} provides two different data samples: 188 GRBs and 31 ``well fit'' GRBs. These ``well fit'' GRBs displayed afterglows that were especially well-modeled by their hydrodynamic simulation, allowing for a high degree of confidence in the $\theta_{obs}/\theta_0$ measurement. All analyses on the data in this paper will be carried out with both data sets.

\subsection{Minimum number of GRBs necessary for model differentiation}
\label{section:minnum}

An important question to consider is \textit{how many} GRBs are necessary in order to confidently distinguish between different models. The following procedure can be carried out in order to find the probability that $N$ GRBs with well-reconstructed $\theta$ are enough to rule out jet structure model A, given that the true underlying distribution is drawn from a probability distribution consistent with jet structure model B:
\begin{enumerate}
\item Generate $N$ random points for measurements of $\theta$ from the probability distribution given by model B (using the Metropolis-Hastings algorithm, for example).
\item Compare the CDF given by model A with the points generated from model B. If the KS test gives a p value below some threshold probability $p_0$, then we can confidently say that these points were not generated by Model A.
\item Repeat steps (i) and (ii) multiple times and record the percentage of times that $N$ random points was enough to exclude model A.
\end{enumerate}

\subsection{Effect of reconstruction precision}

Another important consideration is the effect random noise has on the efficacy of these models. Is it possible that, given $N$ GRBs, we may accidentally rule out the correct model due to un-biased noise in our measurement? The percentage of times that we accidentally rule out the correct distribution can be determined using a similar procedure:
\begin{enumerate}
\item Generate $N$ random points for measurements of $\theta$ from the probability distribution given by your chosen model (using the Metropolis-Hastings algorithm, for example).
\item Add random Gaussian noise (with some standard deviation $\sigma_{noise}$) to each generated point, taking care to not add any noise that would bring the measurement outside of the range $0 \leq \theta \leq 1$.
\item Compare the CDF given by the chosen model (with all free parameters re-fitted to the $N$ noisey points) and see if the KS test gives a p value below $p_0$, ruling out the model.
\item Repeat steps (i), (ii), and (iii) multiple times and record the percentage of times that $N$ random points generated by some underlying distribution appears to have been generated by a different distribution due to some noise of standard deviation $\sigma_{noise}$ applied to the data set.
\end{enumerate}

\section{Results}
\label{results}

\subsection{Constraints from observations}

\begin{figure}
\centering
\includegraphics[width=0.47\textwidth]{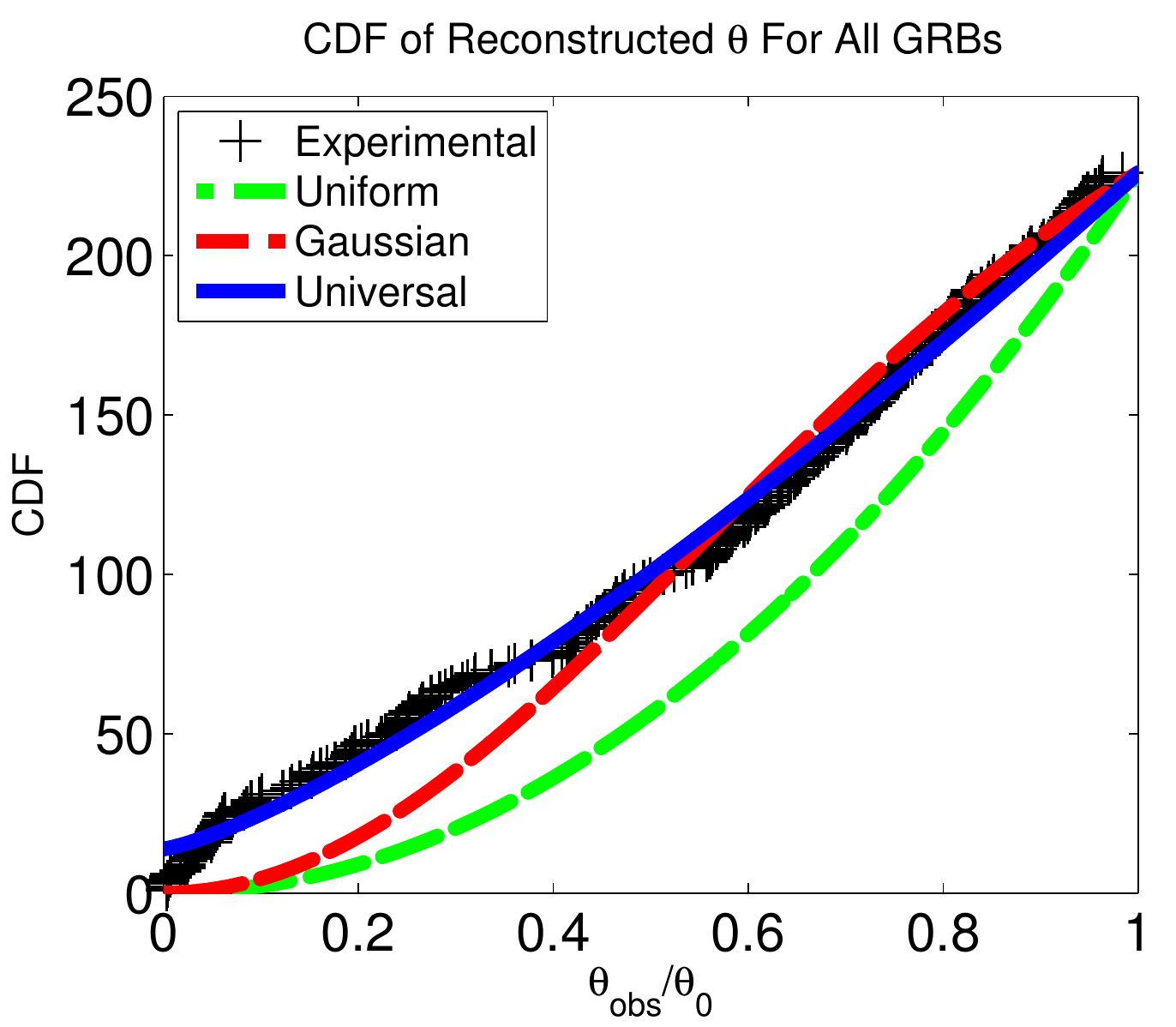}
\includegraphics[width=0.47\textwidth]{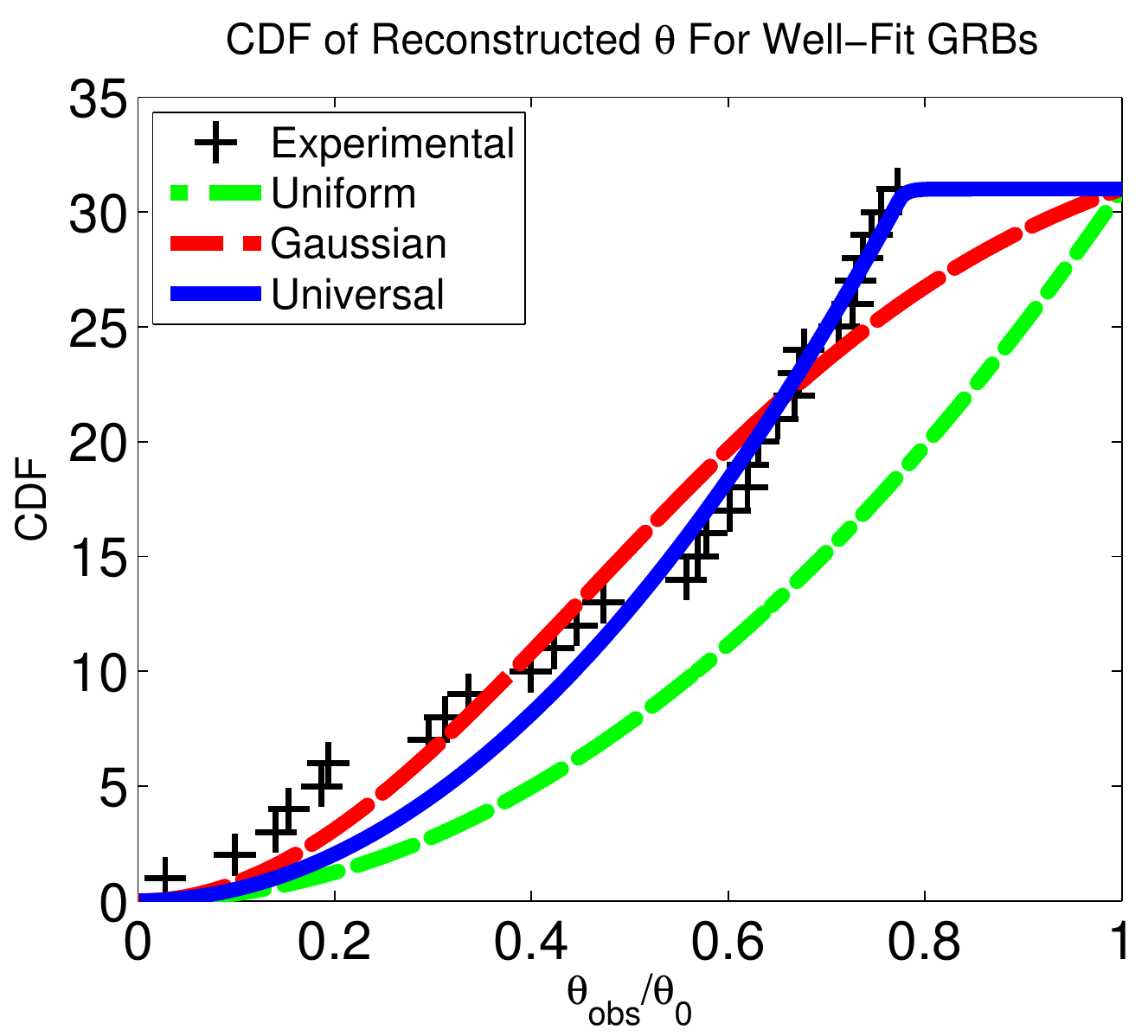}
\caption{Cumulative distribution functions of $\theta = \theta_{obs} / \theta_0$ for the complete (upper) and ``well fit'' (lower) data set from \protect\cite{ryan2015gamma}, along with the expectations from the considered jet structures (see legend) with the best fit parameters.}
\label{fig:cdffits1}
\end{figure}

We compared our three jet structure models with the data sets from \cite{ryan2015gamma}, both ``well fit'' and ``complete'', after recovering the model parameters that best fit the experimental data sets. The CDFs for the data and the best fit models are shown in Fig. \ref{fig:cdffits1} separately for the two datasets. The best fit parameters are shown in Table \ref{tbl:table4}. The obtained p-values using the KS test are listed in Table \ref{tbl:table2}. We find that the uniform structure model is ruled out for both cases at more than $4\sigma$ level. Further, the complete data set is inconsistent with our best fit Gaussian model at a $3\sigma$ level, but is consistent with the ``well fit'' data set. The universal model isn't ruled out in either data set.

\begin{table*}
\small
  \label{tbl:table4}
  \begin{tabular*}{\textwidth}{@{\extracolsep{\fill}}lll}
    \hline
    Model & Parameter fit to all data points & Parameter fit to ``well fit'' data points\\
    \hline
    Gaussian & $\sigma = 0.92$ & $\sigma = 0.63$ \\
    Universal & $\theta_{cut} =  0, s = 0.47$ & $\theta_{cut} = 0.78, s = 74$ \\
    \hline
  \end{tabular*}
  \caption{\ The parameters of the CDF fits using all 226 data points and only the 31 ``well fit'' data points. Note how, while the universal model provides the best fit for both data sets, the full set yields a cut-off angle of 0, while the well-fit set yields a large cut-off angle, past any observed points in the set. This huge cut-off angle and very sharp exponential decay is inconsistent with the implicit assumptions of the model: if the drop-off was actually so steep, with no observed events beyond $\theta_{cut}$, then $\theta_0$ would have been defined to be $\theta_{cut}$, and the uniform model would have fit the data set well. Therefore, the parameter results for the ``well-fit'' set can not be taken to be physical, and rather must be the result of some systematic error.}
\end{table*}

\begin{table*}
\small
  \label{tbl:table2}
  \begin{tabular*}{\textwidth}{@{\extracolsep{\fill}}lll}
    \hline
    Model & p value for all data points & p value for ``well fit'' data points \\
    \hline
    Uniform & $10^{-8}$ & $4 \times 10^{-5}$\\
    Gaussian & $8 \times 10^{-4}$ & 0.3 \\
    Universal & $0.03$ & 0.6 \\
    \hline
  \end{tabular*}
  \caption{\ The p values of the KS test between the experimental $\theta$ distribution and the CDF fits.}
\end{table*}

\subsection{Role of noise}

We now consider the role random noise has in affecting the above comparisons. We use an unbiased noise with standard deviation $\sigma_{noise} = 0.3$ (the typical value for the reconstruction given in \citealt{ryan2015gamma}). We find that the model is unsensitive to noise of this magnitude, as adding such random noise to the data does not affect the obtained constraints, given an underlying probability density arising from either of the structured models. This means, e.g., that the exceptional lack of observed GRBs at large values of $\theta$ in the ``well fit''  dataset must be either the underlying distribution, or the result of systematic errors in the data sample.

\subsection{Minimum number of GRBs necessary for further constraints}

We determined the number of GRBs with well-reconstructed opening and viewing angles needed to constrain the jet structure for different underlying distributions. For this, we first adopt the parameters of the best fits for the different structure models found for the complete and the ``well fit'' data samples. Executing the steps described in Section \ref{section:minnum}, we find that having $\sim 300$ GRBs with accurately reconstructed $\theta$ is sufficient to differentiate between the three jet structure models at $95\%$ confidence level. This number is adequate for the best fit parameters for both datasets considered.

\section{Conclusion}
\label{section:conclusion}

We examined the utility of reconstructed opening and viewing angles for GRBs in determining the underlying jet structure. We considered uniform, Gaussian, and universal jet structures. We used the reconstructed parameters of \cite{ryan2015gamma}, who analyzed over 200 GRBs by fitting their observed afterglow light curve to expected light curves obtained via magnetohydrodynamic simulations.

We find that their results, taken at face value, are inconsistent with a uniform GRB emission profile, and the observations therefore imply a structured jet. Our comparison with Gaussian and universal (power-law) jet profiles are inconclusive; we find that the data cannot rule out either of these models. We find that expected random noise in the reconstructed angles is unlikely to affect this results.

Considering the reconstruction of GRB opening and viewing angles in general, we find that $\sim 300$ well-reconstructed GRBs are likely sufficient to differentiate between the jet structure models we examined. The number of observed GRBs are therefore more than sufficient for an accurate differentiation, if their opening and viewing angles can be determined.

Elucidating the jet structure in GRBs will be important not only for better understanding the mechanism of jet formation, but also in the context of multimessenger observations, for instance in connecting the observed GRB rate with the expected rate of gravitational wave sources (e.g., \cite{2013CQGra..30l3001B}).

Accurate opening and viewing angles for GRBs in the future may require the modification of the method used in \cite{ryan2015gamma}. Ryan et al. simulated a unform jet profile for the comparison with observed outflow light curves. Allowing for alternative jet profiles will be important for quantitative comparisons. Furthermore, understanding systematic uncertainties in the reconstruction needs to be done to ensure that no such uncertainties affect the results.

\vspace{4 mm}
The authors are thankful to Geoffrey Ryan for his help with interpreting the data of \cite{ryan2015gamma}, and to Peter Raffai for providing extensive and detailed comments. This paper was approved for publication by the LIGO Scientific Collaboration. IB and SM are thankful for the generous support of Columbia University in the City of New York. NM is grateful to the I.I. Rabi Scholars Program of Columbia University in the City of New York.

%\bibliographystyle{apj}
%\bibliography{bibliography}

\end{document}